\documentclass{article}

\PassOptionsToPackage{numbers, compress}{natbib}

\usepackage[preprint]{neurips_2023}

\usepackage[utf8]{inputenc} 
\usepackage[T1]{fontenc}    
\usepackage{hyperref}       
\usepackage{url}            
\usepackage{booktabs}       
\usepackage{amsfonts}       
\usepackage{nicefrac}       
\usepackage{microtype}      
\usepackage{xcolor}         
\usepackage{graphicx}
\usepackage{wrapfig}
\usepackage{siunitx}
\usepackage{enumitem}
\usepackage{makecell}  
\usepackage{listings}
\usepackage{float}
\usepackage{caption}  

\title{\textbf{CESPED: a new benchmark for supervised
particle pose estimation in Cryo-EM.}}

\author{%
  Ruben Sanchez-Garcia\(^{1,2}\), Michael Saur\(^2\), Javier Vargas\(^3\), Carl Poelking\(^2\), Charlotte M Deane\(^1\) \And
  \(^1\)\normalfont Department of Statistics, University of Oxford, Oxford OX1 3LB, UK \\
  \texttt{\{ruben.sanchez-garcia, deane\}@stats.ox.ac.uk} \and
  \(^2\)\normalfont Astex Pharmaceuticals, Cambridge CB4 0QA, UK \\
  \texttt{\{michael.saur, carl.poelking\}@astx.com}  \and
  \(^3\)\normalfont Departamento de Optica, Universidad Complutense de Madrid, Madrid 28040, Spain \\
  \texttt{jvargas@fis.ucm.es} \vspace*{-3.5ex} \\
}

\begin{document}

\maketitle

\begin{abstract}

Cryo-EM is a powerful tool for understanding macromolecular structures, yet current methods for structure reconstruction are slow and computationally demanding. To accelerate research on pose estimation, we present CESPED, a new dataset specifically designed for Supervised Pose Estimation in Cryo-EM. Alongside CESPED, we provide a PyTorch package to simplify Cryo-EM data handling and model evaluation. We evaluated the performance of a baseline model, Image2Sphere, on CESPED, which showed promising results but also highlighted the need for further improvements.  Additionally, we illustrate the potential of deep learning-based pose estimators to generalise across different samples,  suggesting a promising path toward more efficient processing strategies.  CESPED is available at https://github.com/oxpig/cesped.
\end{abstract}

\section{Introduction}

\subsection{Cryo-EM Single Particle Analysis}

Determining the structure of macromolecules is crucial to deciphering the intricacies of biological processes and the underlying mechanisms of diseases. With the advent of the resolution revolution, Cryogenic Electron Microscopy (Cryo-EM) has emerged as a leading technique for elucidating structures \citep{Nogales2015a,Callaway2020}. This revolution, driven by significant advances in direct electron detectors and image processing algorithms, has made Cryo-EM a routine, often unrivaled, method for many complex samples \citep{Egelman2016}. Its advantages include, among others, the relative ease of sample preparation compared to other techniques (e.g., x-ray crystallography), the capability to analyze protein complexes previously considered out of reach, and the ability to recover different conformations, offering a dynamic view of molecules in action \citep{Nogales2015}. The pivotal role of Cryo-EM in structural biology was globally recognised in 2017 when the technique was awarded the Nobel Prize in Chemistry.\par

The primary aim of Cryo-EM Single-Particle Analysis (SPA) is to reconstruct the three-dimensional (3D) structure of a given macromolecule at near-atomic resolution, ideally better than 3 Å. This process uses electron beams to capture thousands of two-dimensional  (2D) images of the macromolecules, which are flash frozen in vitreous ice to preserve their native state without the distortions typical of crystalline ice or other fixation methods \citep{Passmore2016}. Each image, called a micrograph, can display several hundred snapshots of the macromolecule (referred to as particle images or just particles) in unknown random orientations. If the orientations of these images were known, the reconstruction task would closely resemble the algorithms used in tomography, which reconstruct 3D volumes from 2D projections taken at predetermined angles \citep{Harauz1986}. However, the unknown orientations of the particles in SPA present a unique challenge not encountered in tomography \citep{Bendory2020}. Compounding this challenge is the inherently low contrast and extremely poor signal-to-noise ratio (SNR) of the images, a consequence of the delicate biological nature of the samples. Given these challenges, a highly sophisticated image processing pipeline is essential to accurately resolve the 3D structure of the macromolecule \citep{Maluenda2019, Tegunov2019}.  \par

The fundamental principle of image processing in SPA is grounded in the intuitive strategy of employing averaging to mitigate noise. Since images are characterised by a low SNR, averaging multiple images of the same particle, assumed to be identical, can significantly enhance the underlying signal \citep{Nogales2015}. However, before averaging can be effectively carried out,  each particle projection must be aligned to a common orientation. This ensures that the differences observed across the images are solely due to noise, allowing its effective cancellation during the averaging process.\par

The standard Cryo-EM image processing pipeline encompasses several key steps, beginning with various preprocessing operations to correct errors such as beam-induced movement blur, followed by particle picking, which extracts the individual particle images from the micrographs \citep{Maluenda2019, Tegunov2019}. Subsequent stages include, among others, clustering (commonly referred to as 2D classification in the context of Cryo-EM) and particle alignment against references, leading to a cleaner subset of the data and an initial low-resolution 3D volume of the protein. This preparatory work sets the stage for the refinement step, a critical phase where the poses of the particles are precisely estimated, a requirement to achieve the high-resolution volumes needed to reveal atomic level details. \par

Traditional refinement algorithms perform pose estimation by exhaustive comparison of experimental particle images and simulated projections of 3D volumes that are iteratively improved \citep{DelaRosa-Trevin2013,Grant2018,Punjani2017, Scheres2012,Tang2007}. When sample homogeneity can be assumed, the simplest approach to the pose estimation problem is the projection matching algorithm \citep{Penczek1994}, which consists of \(T\) iterations of two steps: alignment and reconstruction. First, in the alignment phase, the pose  \((R, s)_i \in \mathrm{SO(3)} \times \mathbb{R}^2\) of each experimental particle image \(x_i\) is set to be the same as the one of the most similar 2D projection of the reference volume \(V^t\) at iteration \(t\),

\begin{equation}\label{traditiona_pose_estimation_1}
(R, s)_i = \arg\min_{{(R, s) \in \mathrm{SO(3)} \times \mathbb{R}^2}} \left\| x_i - f_i \ast P_{(R, s)} V^t \right\|^2
\end{equation}

where \(P_{(R, s)}\) is the projector operator, \(f_i\) is the point spread function of the microscope for the \(i\)-th particle and \(*\) the convolution operator. Then, in the reconstruction phase, a new volume (in reciprocal space) is computed from the estimated poses as 

\begin{equation}\label{traditiona_pose_estimation_2}
\hat{V}^{(t+1)} = \frac{\sum_{i=1}^{N} P_{(R, s)_i}^{-1} \hat{f}_i \hat{x}_i }{ \sum_{i=1}^{N} P_{(R, s)_i}^{-1} \hat{f}_i^2 + C_i}
\end{equation}

with \(\hat{V}\) being the Fourier transform of the volume \(V\), \(C_i\) a constant depending on the SNR, \(\hat{f_i}\) the Fourier transform of the point spread function (CTF, contrast transfer function) and \(N\) the number of particles. This iterative process continues until convergence.
State-of-the-art methods build on this approach, for example, Relion \citep{Scheres2012} employs a Bayesian probabilistic model with a prior for the map, making it much more robust. CryoSPARC \citep{Punjani2017} accelerates Bayesian methods through branch-and-bound search and gradient descent optimisation. See \citep{Bendory2020} for a review.  \par

Despite the innovations aimed at enhancing efficiency, the refinement process still poses significant computational challenges. The primary factor contributing to these challenges is the large number of image comparisons required for each experimental image. Furthermore, the iterative refinement of the volumes, beginning with an initial low-resolution model and progressively improving it, further increases the computational cost, making the refinement stage the most computationally intensive step in the Cryo-EM workflow.

\subsection{Deep Learning for Pose Estimation in Real-World Objects} 

Similar to refinement algorithms in Cryo-EM, traditional pose estimation techniques for real-world images primarily focus on matching 2D images with 3D objects. The significantly higher SNRs characteristic of real-world images enable the use of more sophisticated and efficient methods beyond simple template matching. Among these, landmark-based registration methods are particularly prevalent. Such methods involve extracting distinctive landmarks through various feature extraction techniques \citep{Lowe2004, Bay2008}, followed by a registration process to identify the relative orientation of the landmarks in the image with respect to the reference landmarks \citep{Besl1992} .\par

PoseNet \citep{Kendall2015} was a groundbreaking development in this field, leveraging a Convolutional Neural Network (CNN) to directly regress the absolute pose of an object using quaternions and \(xyz\) shifts. This direct approach contrasts with earlier techniques that relied heavily on feature extraction and landmark identification, allowing for end-to-end pose estimation. Subsequent innovations have built on the foundation laid by PoseNet. Improvements in network architectures \citep{Melekhov2017}, the introduction of more sophisticated loss functions \citep{Kendall2017}, and the incorporation of multitask learning \citep{Xiang2018} have contributed to significant improvements in pose estimation performance.\par

Addressing the inherent challenges of symmetry and occlusion in pose estimation has also seen considerable progress through deep learning. Strategies have evolved from breaking symmetry during the data labelling process \citep{Xiang2014} to implementing loss functions specifically designed to accommodate known symmetries \citep{Xiang2018}.  Probabilistic models offer alternative approaches that either classify poses within a discretised space or explicitly learn the parameters of probability distributions \citep{Deng2022, Mahendran2019, Mohlin2020, Murphy2021, Prokudin2018}.  Due to their probabilistic nature, these models are better suited for challenging datasets with high levels of ambiguity or noise.\par

\subsection{Deep learning methods for Cryo-EM structure determination or pose estimation}

While traditional Cryo-EM refinement algorithms tend to be relatively robust and accurate, they are computationally intensive and slow. In an attempt to overcome this, deep learning (DL) alternatives have begun to emerge. \par

Unsupervised DL methods aim to determine the 3D structure of macromolecules from experimental images alone. Some of them tackle the problem using  a distance learning approach in which the angular distance between pairs of images is estimated as a preprocessing step to retrieve their relative poses \citep{Banjac2021}. Other unsupervised DL methods mirror traditional techniques by maintaining a 3D volume representation to compute 2D projections in a differentiable manner \citep{Donnat2022}. Unlike traditional refinement methods, which compare each experimental particle against all images in an SO(3) projection gallery with up to millions of members, these methods try to limit the number of comparisons between experimental images and projections. For instance, in CryoGAN a 3D volume,  randomly initialised, serves as the generator in the Generative Adversarial Network (GAN) framework \citep{Goodfellow2016}. This generator produces a set of projections from random orientations that are then fed to a discriminator network along with real experimental images. The objective of the training process is to refine the generator until the discriminator can no longer distinguish between the generated projections and the actual experimental images, effectively capturing the underlying 3D structure present in the experimental data. In some other approaches \citep{Levy2022, Levy2022a} ,  particle images are first processed by an encoder designed to predict particle orientations. Following this prediction, a projection of the representation of the volume corresponding to the inferred orientation is rendered. This projection is then directly compared to the original experimental particle image. A loss function is utilized to concurrently refine both the encoder's parameters and the representation of the volume, improving the accuracy of orientation predictions and the fidelity of the reconstructed volume.\par

Supervised DL models, on the other hand, are trained using experimental images and some form of (possibly noisy) labels, such as the poses of prealigned sets of particles. The simplest alternative consists of only an encoder module that predicts the orientation of the particle directly from its image \citep{Jimenez-Moreno2021,Lian2022}. Although supervised approaches offer remarkable efficiency and speed, they require labelled data for training, thus limiting their applicability in \textit{de novo} situations. However, there are use cases where supervised DL methods could offer and advantange. For instance, it should be possible to apply them to on-the-fly pipelines in which a first batch of particles is pre-aligned before the end of the data stream. This initial alignment could be used to train a supervised model to be applied to subsequent batches of data, inferring their poses in real time. Even more interestingly, a pre-trained supervised model could be used to infer poses in different projects, providing the new samples are similar to the training data. This second use case relies on the fact that pose estimation in classical methods is mainly driven by low- to mid-resolution frequencies \citep{Scheres2012a}. As similar proteins have similar low- to mid-resolution frequencies, trained models are expected to generalise to these new samples. In addition, because ligand binding does not generally modify the overall shape of proteins, supervised approaches can be especially valuable in drug discovery, where pre-aligned data for target proteins is often available. \par

In the context of Cryo-EM, only two supervised methods have been proposed to perform direct pose estimation given prealigned particles. DeepAlign \citep{Jimenez-Moreno2021}, a set of CNNs that perform binary classification over a discretisation of \(S^2\), and the approach of \citet{Lian2022} who implemented a CNN to perform direct regression of quaternions. However, due to its limitations, especially for symmetric data, Lian et al. finally adopted a hybrid model with a projector as in some Cryo-EM Unsupervised estimators. Neither of the two methods has been used in practical scenarios.\par

While much slower, classical refinement methods still outperform DL pose estimation models in terms of performance and reliability. This gap can be partly attributed to the unique characteristics of Cryo-EM data, which differ from the natural images DL artchitectures were designed for and, importantly, to the lack of a standardized benchmark that would allow for a direct comparison of methods in order to stimulate progress, much as ImageNet \citep{Deng2009} did for image classification. In this paper, we introduce CESPED (Cryo-EM Supervised Pose Estimation Dataset), a benchmark specifically designed to evaluate supervised pose estimation methods. As the first benchmark dedicated to pose estimation in Cryo-EM, CESPED addresses a crucial gap in the array of available datasets, which have, until now, primarily focused on other Cryo-EM challenges, such as model building \citep{Giri2024} and particle picking \citep{Gyawali2023, Dhakal2023}. CESPED aims to foster advancements in DL methods for particle processing by promoting improvements in supervised pose estimation models, which, due to shared architectural building blocks and data challenges, are likely to benefit methods for related tasks as well.\par

\subsection{Main contributions.} 
In this study, we provide an accessible entry point for a wider scientific audience to engage with the challenges of SPA in Cryo-EM. Toward this goal: 
\begin{itemize}
    \item We compile CESPED (Cryo-EM Supervised Pose Estimation Dataset), an easy-to-use benchmark specifically designed for Supervised Pose Estimation in Cryo-EM.
    \item We implement a PyTorch-based \citep{Paszke2019} package to handle Cryo-EM particle data and to easily compute Cryo-EM quality metrics.
    \item We train and evaluate the Image2Sphere model \citep{Klee2023}, originally developed for real-world pose estimation, on our benchmark, illustrating the utility of our benchmark and shedding light on the transferability of real-world pose estimation models to the Cryo-EM domain.
    \item We present a use case demonstrating that deep learning-based supervised pose estimators have the potential to generalise across related but different samples.
\end{itemize}

\section{Methods}

\subsection{Benchmark Compilation and preprocessing}

In our effort to build a comprehensive benchmark, our primary goal was to identify a diverse set of EMPIAR entries containing at least 200,000 particles, a number deemed sufficient for effective model training. Due to the limitations of EMPIAR's search functionality and the inconsistencies in dataset annotations, we conducted a manual search for entries exceeding this particle count and containing standard Relion files ( .star and .mrcs). Subsequently, we verified the consistency and accuracy of the metadata by running \verb|relion_reconstruct| \citep{Scheres2012} and visually assessing the resulting volumes. This step was crucial for eliminating a significant number of entries due to metadata issues that either crashed the reconstruction process or led to incorrect volumes. To ensure consistent estimation of particle poses, the data was  re-processed using the Relion version 4  auto-refine program \citep{Scheres2012, Kimanius2021} (see Appendix \ref{appx:benchmark_composition} for details). Only entries for which the reconstructed volume exhibited resolution values close to those reported in the literature were selected for inclusion in the benchmark. Finally, for consistency, all images were downsampled to \(1.5\) Å/pixel, with different image dimensions in each entry as macromolecules vary in size. See \ref{appx:benchmark_composition}  for a list of the entries and their properties. \par

The images fed to the deep learning model were preprocessed on-the-fly. We performed per-image normalisation following the standard Cryo-EM procedure, which involves rescaling the intensity so that the background (noise) has a mean of 0 and a standard deviation of 1. We also corrected the contrast inversion caused by the defocus via phase flipping \citep{Penczek2010a}. Finally, since the macromolecule typically represents only between 25\% to 50\% of the whole particle image, the images were cropped so that neighbouring particles are not included.  It is important to note that our benchmark package allows users the flexibility to choose whether or not to apply any of these normalisation steps.\par

The data labels are represented as rotation matrices and then converted into grid indices by finding the closest rotation matrix in the \(\mathrm{SO(3)_{grid}}\). For the cases in which the macromolecule exhibits point symmetry, the labels are expanded as \(L_i={\{g_j  R_i  | g_j \in G}\}\) where \(G\) is the set of rotation matrices given a point symmetry group (e.g., \(C1\)), and \(R_i\) the ground truth rotation matrix. As a result, the labels consist of vectors with \(|G|\) non-zero values and \(|\mathrm{SO(3)_{grid}}| - |G|\) zeros.

\subsection{Baseline model}
We adapted the state-of-the-art Image2Sphere model \citep{Klee2023}. Image2Sphere is a hybrid architecture that uses a ResNet to produce a 2D feature map of the input image, which is then orthographically projected onto a 3D hemisphere and expanded in spherical harmonics. Then, equivariant group convolutions are applied, first with global support on the \(S^2\) sphere, and finally as a refinement step, on SO(3). The output of the model is a probability distribution over a discretised grid of rotation matrices. Other supervised Cryo-EM methods for pose estimation were not considered for this work due to the lack of publicly available code \citep{Lian2022} or their GUI requirements \citep{Jimenez-Moreno2021}.

\subsection{Evaluation metrics}

The most widely used metric in pose estimation is the mean angular error (\(\mathrm{MAnE}\)), averaged across all poses,

\begin{equation}\label{ManE_0}
\mathrm{MAnE} = \frac{1}{N}\sum_{i=1}^{N}  \mathrm{angError}_i  
\end{equation}

The angular error (\(\mathrm{angError}\)) measures the geodesic distance between the predicted and ground truth poses, typically expressed in degrees or radians. This distance can be directly calculated from the rotation matrix of the ground truth pose \(\mathrm{trueR}_i\) and the predicted rotation matrix \(\mathrm{predR}_i\) as

\begin{equation}\label{ManE_01}
\mathrm{angError}_i = \arccos \left(\frac{\mathrm{trace}(\mathrm{trueR}_i \cdot \mathrm{predR}_i^T) - 1}{2} \right)
\end{equation}

When evaluating predicted orientations of macromolecules exhibiting point symmetry, it is necessary to adjust the angular error, as several rotation matrices become equivalent. In this context, the angular error is defined as the minimum geodesic distance between the predicted orientation and any orientation equivalent to the ground truth under the molecule's symmetry group

\begin{equation}\label{ManE_1}
\mathrm{angError}_i = \min_{g_j \in G} \arccos \left(\frac{\mathrm{trace}(g_j \cdot \mathrm{trueR}_i \cdot \mathrm{predR}_i^T) - 1}{2} \right)
\end{equation}

with \(G\) being the set of rotation matrices given a point symmetry group.\par

However, due to the uncertainty in the estimated poses \citep{Jimenez-Moreno2021}, we propose additional metrics. The first one is the confidence-weighted mean-angular-error, 
\begin{equation}\label{wManE_0}
\mathrm{wMAnE} = \frac{ \sum_{i=1}^{N} \mathrm{conf}_i \cdot \mathrm{angError}_i }{ \sum_{i=1}^{N} \mathrm{conf}_i }
\end{equation}

which weights the \( \mathrm{angError_i}\) by \(\mathrm{conf_i}\), the confidence in the ground truth pose, measured as Relion's rlnMaxValueProbDistribution. This confidence estimation is a number between 0 and 1 that measures the probability of the particle having the reported ground truth orientation according to the Relion model. While \( \mathrm{wMAnE}\) is still sensitive to ground truth and confidence estimation errors, due to its simplicity, we used it as the criterion for hyperparameter tuning.\par

The quality of volumes reconstructed from the predicted poses is assessed by comparing them with the ground truth volumes generated from the original poses (see \ref{appx:map_quality_vs_misalignment}). For this comparison, we employ the real space Pearson’s Correlation Coefficient (PCC) and the Fourier Shell Correlation (FSC) Resolution as metrics.

The Pearson’s correlation coefficient is a value between -1 and 1, where values closer to 1 indicate a higher similarity. It measures the linear correlation between the pixels of the two volumes as follows:

\begin{equation}\label{PCC}
\mathrm{PCC}(X,Y) = \frac{\sum_{i=1}^{n}(X_i - \bar{X})(Y_i - \bar{Y})}{\sqrt{\sum_{i=1}^{n}(X_i - \bar{X})^2} \sqrt{\sum_{i=1}^{n}(Y_i - \bar{Y})^2}} \quad
\end{equation}

with \(X_i\) and \(Y_i\) being the pixel \(i\) of the two volumes, \(n\) the number of voxels, and \(\bar{X}\) and \(\bar{Y}\) the average value of the volumes. \par

The \(\text{FSC}\) quantifies the correlation between two signals at different spatial frequencies. For each frequency \(k\), a value between -1 and 1 (with higher values indicating greater similarity) is computed by comparing the concentric shells in the Fourier transforms of the two volumes corresponding to \(k\):

\begin{equation}\label{FSC}
\text{FSC}_k(X,Y) = \frac{\sum_{\mathbf{r} \in \text{shell}(k)} \hat{X}(\mathbf{r}) \cdot \hat{Y}^*(\mathbf{r})}{\sqrt{\left(\sum_{\mathbf{r} \in \text{shell}(k)} |\hat{X}(\mathbf{r})|^2\right) \cdot \left(\sum_{\mathbf{r} \in \text{shell}(k)} F|\hat{Y}(\mathbf{r})|^2\right)}}
\end{equation}

where $\hat{X}(\mathbf{r})$ and $\hat{Y}(\mathbf{r})$ represent the Fourier transforms of the two volumes at frequency $\mathbf{r}$, $\text{shell}(k)$ is the shell of frequency $k$ and $\hat{Y}^*(\mathbf{r})$ is the complex conjugate of $\hat{Y}(\mathbf{r})$. 

To summarize the FSC curves into a single number, the FSC Resolution is computed by selecting a threshold \(t\) and identifying the highest frequency \(k\) such that \(\text{FSC}_{k}(X,Y) < t\). As thresholds we employ the commonly used values of 0.5 (\(\mathrm{FSCR_{0.5}}\)) and 0.143 (\(\mathrm{FSCR_{0.143}}\)), which corresponds to the highest frequency at which the two maps agree with an SNR of 1 and 0.5 respectively \citep{Rosenthal2003}. 

In order to decouple the different quality levels of the different benchmark entries, we report the differences of the metrics with respect to the ground truth levels, estimated from the half-maps of the ground truth, 

\begin{equation}\label{delta_pcc}
    \Delta\mathrm{PCC} = \mathrm{PCC}(GT_0,GT_1) - \mathrm{PCC}(GT, V)
\end{equation}
and
\begin{equation}\label{delta_fsc}
    \Delta\mathrm{FSC} = \mathrm{FSC}(GT_0,GT_1) - \mathrm{FSC}(GT, V)
\end{equation}
where \(GT\) is the ground truth map, \(GT_i\) is the ground truth map reconstructed with the \(i\)-th half of the data, and \(V\) is the 3D volume reconstructed with the predicted poses (see Figure \ref{fig:eval_protocol}).

\subsection{Training}
Each benchmark entry was trained independently with the same hyperparameters (see \ref{appx:hparams}). Due to the uncertainty in the estimated orientations, we employed a weighted cross-entropy loss using the pose reliability estimate of each particle as the per-image weight 

\begin{equation}
  L=\frac{1}{N} \sum_{i=1}^{N} \sum_{c=1}^{C} - \mathrm{conf}_i \cdot P(R_{c,i}) \log(Q(R_{c,i}))  
\end{equation}

 where \(Q(R_{c,i})\) is the predicted probability for the rotation matrix with grid index \(c\) and \(P(R_{c,i})\) is \(1/|G|\) when any of the ground truth matrices is \(c\) and zero otherwise.

\subsection{Evaluation protocol}

\begin{wrapfigure}{r}{0.5\textwidth} 
  \centering
  \includegraphics[scale=0.4]{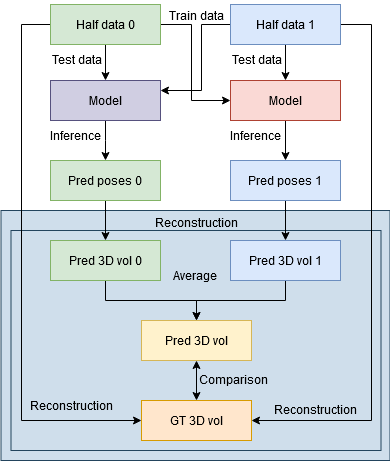}
  \caption{Evaluation protocol inspired by the Cryo-EM Gold Standard.  For each entry, the dataset is randomly split into two subsets (Half data 0 and Half data 1) that are processed independently.  Then, each half of the data is used to  train a different model that will be used to infer the poses of the other half of the data. From the inferred poses, two reconstructed volumes can be obtained, one for each half of the data. The two reconstructed volumes can be combined and compared to the ground truth volume, which is obtained from the ground truth poses. The grey box represents the automatic evaluation tool, that takes as input the predictions for both data half sets and internally computes the required reconstructed volumes to perform the comparisons.}
  \label{fig:eval_protocol}
  
\end{wrapfigure}

Due to the uncertainty in the ground truth labels and the fact that what matters to Cryo-EM practitioners is the quality of the reconstructed volume, we devised an evaluation protocol inspired by the Cryo-EM gold standard \citep{Henderson2012}, which is a per-entry 2-fold cross-validation strategy in which the poses of each half of the data are independently estimated and used to reconstruct two volumes (half-maps). For benchmarking supervised methods, it involves training an independent model for each half of the dataset to infer the poses of the other half of the dataset. After that, the final 3D volume is computed by reconstructing the two half-maps and averaging them. The final averaged map can then be compared with the ground truth map obtained from the original poses (see Figure \ref{fig:eval_protocol}).  It is important to note that the FSC resolution values derived from this comparison are analogous to map-to-model FSC resolution estimations and not equivalent to the gold standard half-to-half resolution.\par

Since Image2Sphere predicts only rotation matrices but not image shifts, when reconstructing the volumes, we employed the ground truth translations. This could result in an overoptimistic estimation of performance, however, since the effect of the translations is tightly coupled with the accuracy of the angular estimation, this overestimation should be small. We leave for future work the full inference of both the rotations and translations. Finally, to avoid overfitting to the validation set, we performed hyperparameter tuning only one half of the data using \(\mathrm{wMAnE}\) as a metric.

\section{Results and discussion}
\subsection{Benchmark, ParticlesDataset class and evaluation tool}
Our benchmark consists of a diverse set of eight macromolecules, with an average number of ~300K particles including soluble and membrane macromolecules, symmetric and asymmetric complexes, and resolutions ranging from \SI{5}{\angstrom} to \SI{3.2}{\angstrom} (see \ref{appx:benchmark_composition}). For each particle in the dataset, we provide its image and estimated pose together with an estimate of the reliability of the poses. The benchmark can be automatically downloaded from  Zenodo \citep{Zenodo2013} using our \(\mathrm{cesped}\) Python package.\par

The package includes a \verb|ParticlesDataset| class, which implements the PyTorch \verb|Dataset| API for seamless integration. It also offers optional yet recommended preprocessing steps commonly adopted in Cryo-EM (e.g., image normalisation, phase flipping), and specialised data augmentation techniques, like affine transformations that adjust both the image and its corresponding pose (see \ref{appx:hparams}). While the \(\mathrm{cesped}\) package was designed with PyTorch in mind, the benchmark is accessible to a broader audience, as the data is stored in standard formats and accompanied by utility programs to assist users of other frameworks in adopting the CESPED benchmark.\par

Additionally, the package offers an automatic evaluation pipeline that only requires as inputs the predicted poses (grey box in Figure \ref{fig:eval_protocol}). For ease of use, a Singularity\footnote{https://zenodo.org/records/4667718} image definition file is included, eliminating the need to install Cryo-EM-specific software like Relion. This design enables those without Cryo-EM experience to utilize the \verb|cesped| benchmark and package as effortlessly as they would with standard datasets such as MNIST. Usage examples can be found in \ref{appx:usage}.

\subsection{Performance of the baseline model on the benchmark}

\begin{figure}[ht]
    \centering
    \includegraphics[width=1\linewidth]{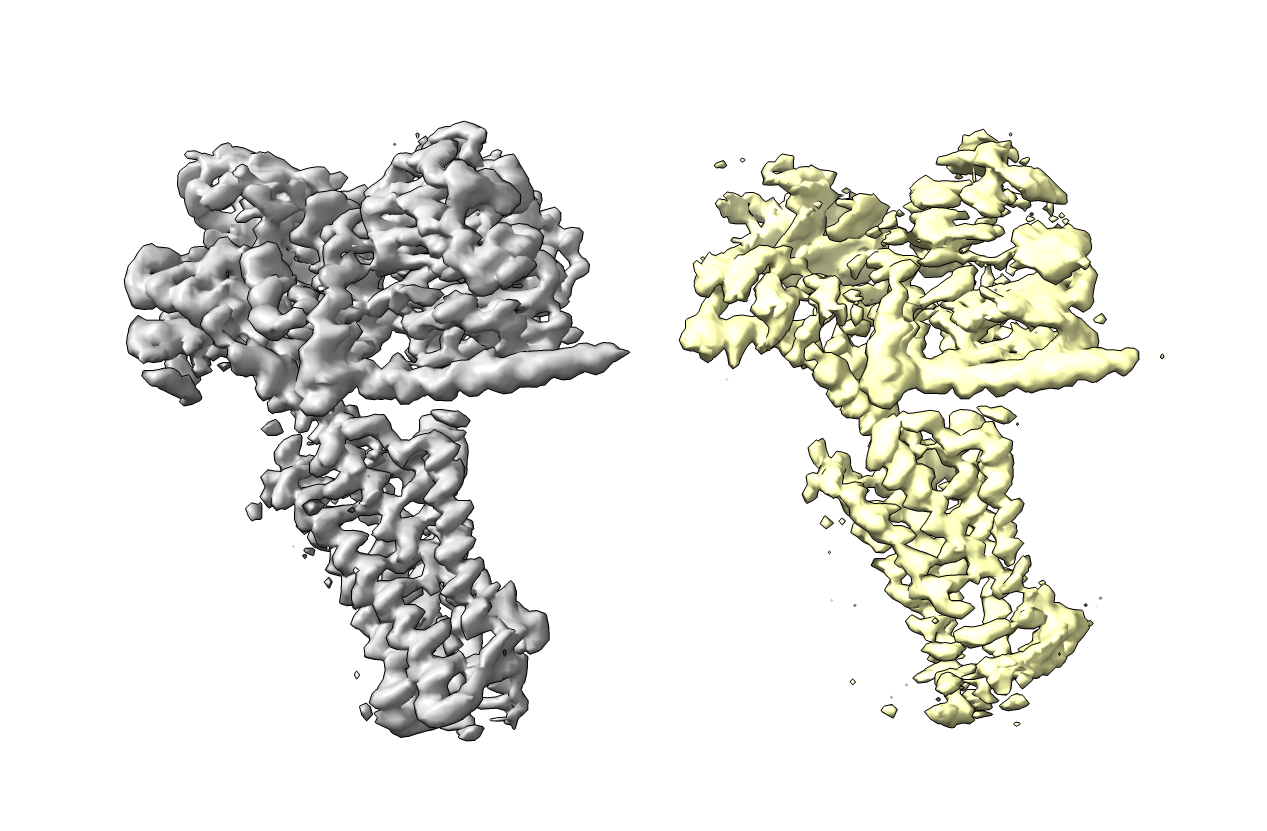}
    \caption{Ground truth reconstruction for EMPIAR-10786 (grey) and reconstruction using the angles predicted with the Image2Sphere model (yellow).}
    \label{fig:empiar-10786-vol}
\end{figure}

Table \ref{Image2Sphere-benchmark-results} summarises the results of the Image2Sphere \citep{Klee2023} model on our benchmark, with per-entry results in \ref{appx:per_entry_results}. While the \(\mathrm{wMAnE}\)  is \( \sim \) 24°, for the best cases, the error is as small as 9°. The \(\Delta\mathrm{PCC}\)  for the worse cases is > 0.1, highlighting that, for some entries, the reconstructed volumes are far from the ground truth solution. For a few cases, the results are much better, with \(\Delta\mathrm{PCC}\) < 0.03. In terms of prediction vs ground truth \(\mathrm{FSCR_{0.5}}\), most maps are in the 8-6 Å range, with \(\Delta\mathrm{FSCR_{0.5}}\)  of 3.6 Å. However, the \(\mathrm{FSCR_{0.143}}\) values between 4-5 Å, indicate better correlation at lower signal levels. This visually translates into a relatively well-resolved central part of the map that becomes blurrier away from the centre (see Figure \ref{fig:empiar-10786-vol} and 
\ref{appx:Reconstructed_volumes}). For the top-performing cases, a simple and fast local refinement of the predicted poses is sufficient to obtain high-resolution reconstructions comparable to ground truth volumes, at a computational cost threefold less than global refinement (\ref{appx:Local_refinement}). Since the Image2Sphere model inference takes only minutes, far less than the hours needed for traditional refinement, further improvements could reduce computing times by at least one order of magnitude if local refinement is no longer needed (see \ref{appx:Running_times} for running times).

\begin{table}[H]
  \caption{Image2Sphere results on CESPED. \(\mathrm{MAnE}\) and \(\mathrm{wMAnE}\) measure angular errors in predicted poses. \(\Delta\mathrm{PCC}\) and  \(\Delta\mathrm{FSCR}\)  measure the reduction in quality of the predicted volumes compared to the ground truth. The mean and std of the metrics are computed over the seven benchmark entries.}
  \label{Image2Sphere-benchmark-results}
  \centering
  \begin{tabular}{lcccccc}  
    \toprule
          & \(\mathrm{MAnE}\) (°)  & \(\mathrm{wMAnE}\) (°) & \(\Delta\mathrm{PCC}\) &  \(\Delta\mathrm{FSCR_{0.5}}\) (Å) &  \(\Delta\mathrm{FSCR_{0.143}}\) (Å) \\
    \midrule
    mean (std) & 28.7 (12.7) & 23.8 (12.2) & 0.059 (0.033) & 3.4 (0.6) & 1.3 (0.7) \\
    \bottomrule
  \end{tabular}
\end{table}

Given the inherent difficulties of Cryo-EM data, the fact that a generic pose estimation model can produce meaningful results in some examples without major modifications suggests that equivariant architectures can be useful for the Cryo-EM data domain.

\subsection{Example of model generalisability across samples }

One of the main potential applications of Supervised Pose Estimation models is to infer poses on similar, yet different projects. In this section, we illustrate this use case by using an Image2Sphere model trained on the EMPIAR-10280 dataset, to predict poses of the same protein under different experimental conditions (EMPIAR-10278 dataset).
 
\begin{figure}[H]
    \centering
    \includegraphics[width=1\linewidth]{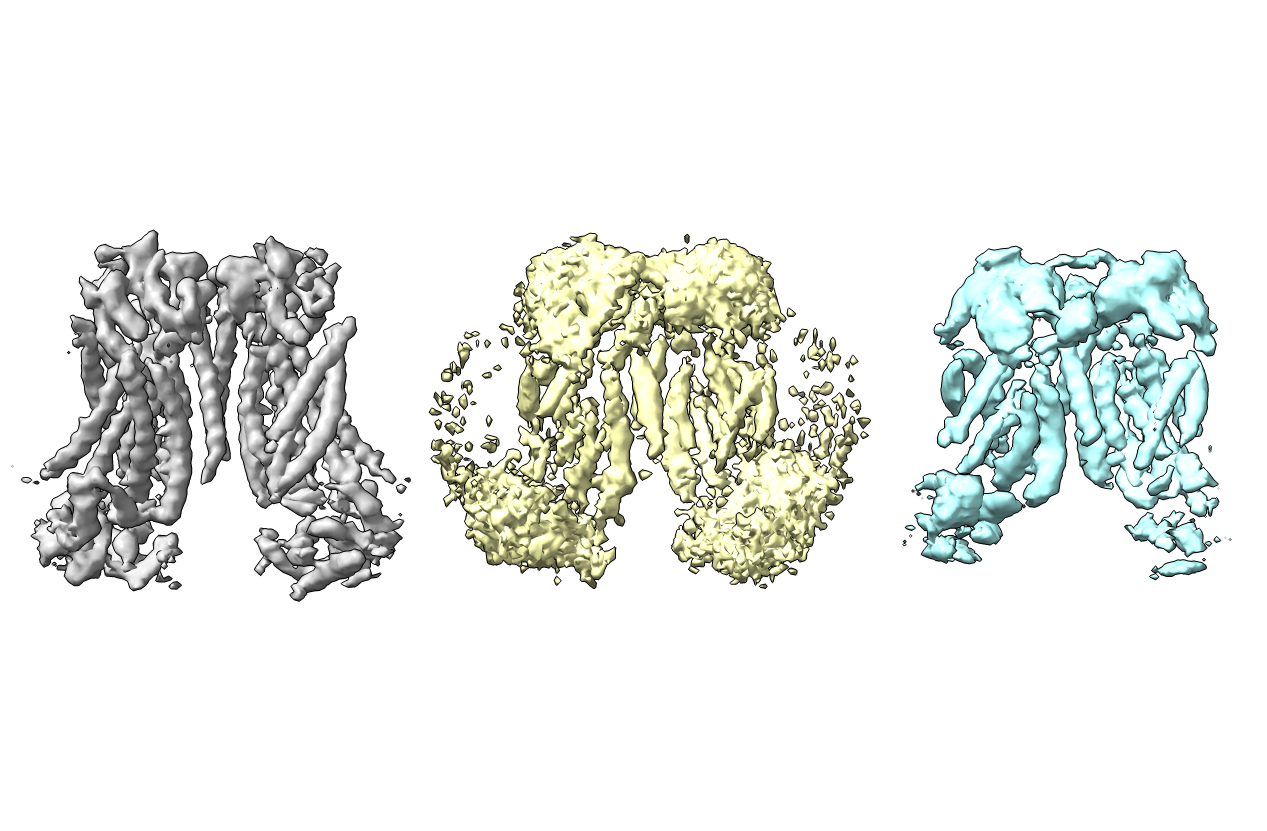}
    \caption{
Generalisability use case. Left and centre, reconstructions for the EMPIAR-10278 dataset from the ground truth poses (grey) and the predicted poses (yellow) using a model trained on the EMPIAR-10280 data. Right, reconstruction for the EMPIAR-10280 dataset using a model trained on the EMPIAR-10280 data used as control (cyan). The model used for the yellow and cyan volumes is the same, but the particles fed to the model come from different datasets of the same protein in different conditions. Although of lower quality than the cyan volume, the yellow volume demonstrates similarities to the ground truth map (grey).}
    \label{fig:_generalizability_10280-vol}
\end{figure}

Figure \ref{fig:_generalizability_10280-vol} showcases three reconstructed volumes: (1) EMPIAR-10278 using ground truth poses (grey); (2) EMPIAR-10278 with poses predicted by the model trained on EMPIAR-10280 (yellow), illustrating the model's generalisability; and (3) EMPIAR-10280 using poses inferred by the model trained on its own dataset, serving as a control for model performance. As expected, the EMPIAR-10278 map reconstructed with original poses shows superior quality compared to the others. Similarly, the EMPIAR-10280 map generated from the model trained on EMPIAR-10280 exhibits better quality than the EMPIAR-10278 map inferred using the EMPIAR-10280 model, reflecting the differences between the two datasets despite containing the same protein. Independently of these quality differences, the model's capacity for generalization across datasets is evident through visual inspection of the EMPIAR-10278 inferred map (yellow), as the overall shape of the protein and several key secondary structure elements are clearly recognizable. This suggests that further improvements in the model could lead to the desired goal of training the model once and then inferring the poses of similar datasets at much faster speeds.

\subsection{Challenges and future directions}
Cryo-EM particle images are fundamentally different from the kinds of images encountered in other fields. One of the most critical challenges is their poor SNR, which can be as low as $0.01$ \citep{Bendory2020}. While some methods have tried to mitigate this issue by applying filtering techniques \citep{Levy2022a} or using CNNs with larger kernel sizes \citep{Bepler2019, Jimenez-Moreno2021, Sanchez-Garcia2018}, these solutions are not entirely effective. \par

Symmetry presents another complex facet of Cryo-EM data. Exploiting symmetry can drastically reduce the computational requirements for pose estimation, but it can also prevent simple models from learning. The unique combination of rotationally equivariant convolutions with the probabilistic estimation of poses makes the Image2Sphere model an ideal candidate to exploit this feature. However, the hybrid \(S^2\) / SO(3) formalism means that the separation of rotational degrees of freedom from translational in-plane shifts is not easily achieved within this framework. A significant area for future work lies in leveraging rotational equivariance and translational equivariance for the joint estimation of the rotational and translational components of the poses (e.g. SE(3) equivariance). \par

In this work we have considered only the case of homogeneous refinement, which assumes that all particles are projections from a single macromolecule in a unique conformation. However, this is not always the case and our benchmark could potentially be extended to deal with such examples. Models would then need to perform conformation classification alongside pose estimation.

\section{Conclusions}
Pose estimation is one of the most critical steps of the Cryo-EM processing pipeline, and while current algorithms are relatively robust and reliable, they are also computationally slow. Deep learning holds the promise of overcoming these challenges, but achieving this potential hinges on improvements in accuracy and reliability, for which systematic benchmarking is required. In this study, we introduce a benchmark specifically designed for Supervised Pose Inference of Cryo-EM particles, along with a suite of code utilities to assist machine learning practitioners unfamiliar with Cryo-EM. We also present a real-world image pose prediction model applied to our benchmark, demonstrating promising preliminary results on a subset of the data. This preliminary success suggests that addressing Cryo-EM-specific challenges, such as high noise levels and label inaccuracies, could lead to even better performance. The improvements in models for this benchmark will not only pave the way for more effective Supervised Pose Prediction models, but are also likely to give rise to innovative approaches to closely related challenges like Unsupervised Pose Estimation and Heterogeneity Analysis. Ultimately, those advancements could serve as a catalyst for even further developments, leading to a new paradigm in Cryo-EM image processing.

\section{Availability}
CESPED dataset and code can be found at https://github.com/oxpig/cesped.

\section{Acknowledgments}
Ruben Sanchez-Garcia is funded by an Astex Pharmaceuticals Sustaining Innovation Post-Doctoral Award. Javier Vargas is financially supported by the Spanish Ministerio de Ciencia e Innovación, Grant PID2022-137548OB-I00 funded by MCIN/AEI/10.13039/501100011033/. 

\bibliography{mendeley}
\bibliographystyle{plainnat}  

\newpage
\setcounter{table}{0} 
\setcounter{figure}{0} 

\appendix
\renewcommand{\thesection}{Appendix \Alph{section}}
\section{Benchmark composition}\label{appx:benchmark_composition}

\begin{table}[h]  
\captionsetup{name=Supplementary Table, labelsep=period} 
\caption{CESPED benchmark entries.}  
\small  
\centering  
\begin{tabular}{| l | p{4cm} | l | l | l | l | l |}  
\hline
\rotatebox{90}{\textbf{EMPIAR ID}} & \textbf{Composition} & \rotatebox{90}{\textbf{Symmetry}} & \rotatebox{90}{\textbf{Image pixels}} & \rotatebox{90}{\textbf{FSCR\textsubscript{0.143} (Å)}} & \rotatebox{90}{\makecell{\textbf{Masked} \\ \textbf{FSCR\textsubscript{0.143} (Å)}}} & \rotatebox{90}{\textbf{\# particles}} \\
\hline
10166 & Human 26S proteasome bound to the
  chemotherapeutic Oprozomib & C1 & 284 & 5.0 & 3.9 & 238631 \\
\hline
10786 & Substance P-Neurokinin Receptor G
  protein complexes (SP-NK1R-miniGs399) & C1 & 184 & 3.3 & 3.0* & 288659 \\
\hline
10280 & Calcium-bound TMEM16F in nanodisc with
  supplement of PIP2 & C2 & 182 & 3.6 & 3.0* & 459504 \\
\hline
11120 & M22 bound TSHR Gs 7TM G protein& C1 & 232 & 3.4 & 3.0* & 244973 \\
\hline
10409 & Replicating SARS-CoV-2 polymerase (Map 1) & C1 & 240 & 3.3 & 3.0* & 406001 \\
\hline
10374 & Human ABCG2 transporter with inhibitor MZ29 and 5D3-Fab
    & C2 & 216 & 3.7 & 3.0* & 323681 \\
\hline
10399 & Arabinofuranosyltransferase AftD 
from Mycobacteria & C1 & 184 & 3.2& 3.1 & 490616 \\
\hline
10648 & PKM2 in complex with Compound 5 & D2 & 222 & 3.7 & 3.3 & 234956 \\
\hline
\hline 
simulated 10648 & Same PKM2 dataset as in 10648, but with simulated images & D2 & 222 & 3.5 & 3.4 & 138848 \\
\hline
consensus 10648 & Same PKM2 dataset as in 10648, but with consensus angles & D2 & 222 & 3.8 & 3.4 & 234956 \\
\hline
\end{tabular}

\begin{itemize}
    \item * Nyquist Frequency at 1.5 Å/pixel; Resolution is estimated at the usual threshold 0.143.
    \item Reported FSCR\textsubscript{0.143} values were obtained directly from the relion\_refine logs while Masked FSCR\textsubscript{0.143} values were collected from the relion\_postprocess logs.
\end{itemize}
\end{table}

Particle poses were estimated using the Relion version 4 auto-refine program\citep{Scheres2012, Kimanius2021}. As starting model we used the map obtained with:

\verb|relion_reconstruct --pad 2.0 --ctf --i original_poses.star|  
 \par \verb|--sym $SYMMETRY|
 \verb| --o reconstructd_map.mrc| \par 

The mask was created using: \par 

\verb|relion_mask_create --i reconstructed_map.mrc --o mask.mrc --lowpass  15.0 |
\par\verb|-extend_inimask 3 --width_soft_edge 6 --ini_threshold $THRESHOLD| \par
with \verb|$THRESHOLD| manually selected for each entry.

The auto-refine command used was \par

\verb|mpirun -np 5 relion_refine_mpi --i original_poses.star --particle_diameter | 
\par\verb|$DIAMETER --ctf  --zero_mask  --firstiter_cc  --ini_high 40.0 | 
\par\verb|--sym $SYMMETRY --ref reconstructd_map.mrc --norm  --scale  |\par \verb|--solvent_mask mask.mrc --o outputdir/run --oversampling 1 --flatten_solvent| 
\par\verb|--solvent_correct_fsc  --pad 2 --auto_local_healpix_order 4 --healpix_order 2 |
\par\verb|--offset_range 5.0 --offset_step 2.0 --auto_refine  --split_random_halves |
\par\verb| --low_resol_join_halves 40 --dont_combine_weights_via_disc|

The simulated dataset was generated with the following command:\par
\verb| relion_project --i original_poses.star --ang original_poses.star | \par \verb|  --ang_simulate original_poses.star --o simulated_dir/simulated | \par \verb|  --simulate --adjust_simulation_SNR 2.0 --ctf |

The consensus dataset was generated using the compare angles protocol from Scipion Xmipp \citep{Sorzano2022, Jimenez-Moreno2021}, incorporating both our original Relion refinement output and a refinement performed with cisTEM \citep{Grant2018}. An angular distance threshold of 5° was employed.

\section{Image2Sphere and training hyperparameters} \label{appx:hparams}

Our Image2Sphere model follows the implementation of \citet{Klee2023} with the following configuration:

\begin{itemize}
    \item Feature extractor: ResNet152 \citep{He2016} with default parameters as implemented in torchvision using imageNet weights. The input images are resized to 256 pixels before being fed, giving a feature map of shape 2048x8x8. Since the input images only contain one channel, but the ResNet expects 3 channels, two additional channels were added by applying a Gaussian filter with sigma 1 and 2 to the input image.
    \item Image projector to S2: Default orthographic projector with HEALPix \citep{Gorski2004} grid  order 3 (\(\sim7.5^\circ\)), where only 50\% of the grid points are sampled. The feature map is projected from 2048 channels to 512 using a 1x1 Conv2d and then converted to spherical harmonics with \(l_{max}=8\).
    \item S2 convolution: 512 filters with global support on a HEALPix grid of order 3.
    \item SO(3) convolution: 16 filters with local support (max\_beta= \(\pi/8\), max\_gamma=\(2\pi\), n\_alpha=8, n\_beta=3).
    \item Probability distribution discretization: HEALPix grid of order 4 (\(\sim3.7^\circ\)).
\end{itemize}

Training was conducted using RAdam \citep{Liu2019} as optimizer with an initial
learning rate of 1e-3. A weight decay of 1e-5 was employed. The learning rate
was halved each time the validation loss stagnated during 10 epochs. The
training was stopped when the number of epochs reached 400 or the validation
loss did not improve for 12 epochs.

Data augmentation was conducted with the following composed
transformations:
\begin{itemize}
    \item Random shift from -5\% to 5\% with probability 0.5.
    \item Random rotation from -20° to 20° with probability 0.5.
    \item Random 90° rotation with probability 1.
    \item Uniform noise addition with a random scale from 0 to 2 with probability 0.2.
    \item Gaussian noise addition with a random standard deviation from 0 to 0.5 with probability 0.2.
    \item Random zoom-in of size 0\% to 5\% with probability 0.2.
    \item Random erasing of patches of size 0\% to 2\% with probability 0.1.
\end{itemize}
Notice that rotation transformations require adjustments in the
ground truth labels.

\section{cesped package usage example}\label{appx:usage}

\definecolor{codegreen}{rgb}{0,0.6,0}
\definecolor{codegray}{rgb}{0.5,0.5,0.5}
\definecolor{codepurple}{rgb}{0.58,0,0.82}
\definecolor{backcolour}{rgb}{0.95,0.95,0.92}

\lstdefinestyle{mystyle}{
    backgroundcolor=\color{backcolour},   
    commentstyle=\color{codegreen},
    keywordstyle=\color{magenta},
    numberstyle=\tiny\color{codegray},
    stringstyle=\color{codepurple},
    basicstyle=\ttfamily\footnotesize,
    breakatwhitespace=false,         
    breaklines=true,                 
    captionpos=b,                    
    keepspaces=true,                 
    numbers=left,                    
    numbersep=5pt,                  
    showspaces=false,                
    showstringspaces=false,
    showtabs=false,                  
    tabsize=2
}

\lstset{style=mystyle}

Dataset instantiation only requires providing the name of the target (a string like \verb|"10280"|) and the half-set number (0 or 1) (Listing \ref{training_loop_code}). \verb|ParticlesDataset|s can be directly used as datasets in PyTorch \verb|DataLoader|(s).

\begin{lstlisting}[language=Python,label=training_loop_code,caption=Example of how to load and use a CESPED benchmark entry in a training loop]
import torch
from cesped.particlesDataset import ParticlesDataset

listOfEntries = ParticlesDataset.getCESPEDEntries()
targetName, halfset = listOfEntries[0] #We will work with the first example
dataset = ParticlesDataset(targetName, halfset)
dl = DataLoader(dataset, batch_size=32, num_workers=4)
for batch in dl:
    iid, img, (rotMat, xyShiftAngs, confidence), metadata = batch

    #iid is the id of the particle (a string)
    #img is a batch of Bx1xNxN images
    #rotMat is a batch of rotation matrices Bx3x3
    #xyShiftAngs is a batch of image shifts in Angstroms Bx2
    #confidence is a batch of numbers between 0 and 1, Bx1
    #metata is a dict of names:values with particle information 
    
    predRot = model(img)
    loss = loss_function(predRot, rotMat)
    loss.backward()
    optimizer.step()
    optimizer.zero_grad()

\end{lstlisting}

\verb|ParticlesDataset| objects can also be used to update the metadata with newly predicted poses and to save the results in Relion star format, commonly used in Cryo-EM software (Listing \ref{write_results_loop_code}).

\begin{lstlisting}[language=Python,label=write_results_loop_code,caption=Example of how to save predictions for usage in Cryo-EM packages and evaluation]
for iid, pred_rotmats, maxprob in predictions:
    #iid is the list of ids of the particles (string)
    #pred_rotmats is a batch of predicted rotation matrices Bx3x3
    #maxprob is a batch of numbers, between 0 and 1, Bx1
    #that indicates the confidence in the prediction (e.g., softmax values)
    n_preds = pred_rotmats.shape[0]
    dataset.updateMd(ids=iid, angles=pred_rotmats,
                     shifts=torch.zeros(n_preds, 2), 
                     confidence=maxprob,
                     angles_format="rotmat")
dataset.saveMd("predictions.star") #Save dataset as an starfile
\end{lstlisting}

Once the predictions are computed for the two halves of the benchmark entry, evaluation can be automatically computed by providing the starfiles of both predictions via a command line tool(Listing \ref{evaluate_script}) or a function. While you can use your local installation of Relion, we also provide a Singularity definition file so that you do not need to manually install it. See instructions at https://github.com/rsanchezgarc/cesped,

\begin{lstlisting}[language=bash,label=evaluate_script,caption=Evaluation script execution]
python -m cesped.evaluateEntry  --predictionType SO3 --targetName 10280 
--half0PredsFname particles_preds_0.star 
--half1PredsFname particles_preds_1.star 
--n_cpus 12 --outdir evaluation/
\end{lstlisting}
\begin{lstlisting}[language=Python,label=evaluate_function,caption=Evaluation function example]

from cesped.evaluateEntry import evaluate
evaluation_metrics = evaluate(targetName="10280",
            half0PredsFname="particles_preds_0.star", 
            half1PredsFname="particles_preds_1.star", 
            predictionType="SO3", #Literal["S2", "SO3", "SO3xR2"], 
            usePredConfidence=True,
            n_cpus=4, 
            outdir="output/directory")

\end{lstlisting}

\pagebreak

\section{Image2Sphere per-entry results}\label{appx:per_entry_results}

This section contains per-entry statistics for the Image2Sphere model predictions using the evaluation protocol proposed in the main text. The last two rows correspond to different versions of the 10648 entry and have not been included in the Table \ref{Image2Sphere-benchmark-results}. In addition to angular error measurements, the other metrics compare the ground truth map (\(\mathrm{GT}\)) against the map reconstructed from the predicted poses (\(\mathrm{V}\)), namely the \(\mathrm{PCC}(GT,V)\) and \(\mathrm{FSCR_{t}}(GT,V)\), where $t$ denotes the threshold 0.5 or 0.143, where reported. GT is obtained by employing \verb|relion_reconstruct| on the ground truth poses (that were estimated with \verb|relion_refine --auto_refine|).  The reconstructed map V is generated with \verb|relion_reconstruct|  from the predicted poses. \par
In addition, we also report half-to-half map metrics, which are commonly employed in traditional Cryo-EM algorithms and in Unsupervised DL methods and can be used to compare them to Supervised DL methods. In particular, we compute \(\mathrm{PCC}(GT_0,GT_1)\), \(\mathrm{PCC}(V_0,V_1)\), \(\mathrm{FSCR_{t}}(GT_0,GT_1)\) and \(\mathrm{FSCR_{t}}(V_0,V_1)\), where 0 and 1 denote the dataset half.  Thus  \(V_0\) is the map reconstructed from the predicted poses of the half dataset 0 using a model trained on the dataset 1. \(GT_0\) is obtained as GT, but using only the ground truth poses of the half dataset 0. 
\begin{table}[h]  
\captionsetup{name=Supplementary Table, labelsep=period} 
\caption{Per-entry CESPED benchmark results using an Image2Sphere model.}  
\small  
\centering  
\begin{tabular}{| l | l | l | l | l | l | l | l | l | l | l | l |} 
\hline
\rotatebox{90}{\textbf{EMPIAR ID}} & \rotatebox{90}{\textbf{\(\mathrm{MAnE}\) (°)}} & \rotatebox{90}{\textbf{\(\mathrm{wMAnE}\) (°)}} & \rotatebox{90}{ \(\mathrm{PCC(V_0,V_1)}\)} & \rotatebox{90}{\textbf{\(\mathrm{PCC}(GT, V)\)}} & \rotatebox{90}{\makecell{\(\mathrm{FSCR_{0.143}(V_0,V_1)}\) \\ (Å)}} & \rotatebox{90}{\makecell{\(\mathrm{FSCR_{0.5}(V_0,V_1)}\) \\  (Å)}} 
& \rotatebox{90}{\makecell{\(\mathrm{FSCR_{0.143}(GT, V)}\)\\  (Å)}} & \rotatebox{90}{\makecell{\(\mathrm{FSCR_{0.5}}(GT, V)\) \\  (Å)}}  & \rotatebox{90}{\makecell{{\(\mathrm{FSCR_{0.143}(GT_0,GT_1)}\) } \\ (Å)}} & \rotatebox{90}{\makecell{{\(\mathrm{FSCR_{0.5}(GT_0,GT_1)}\)} \\  (Å)}}  & \rotatebox{90}{\textbf{\(\mathrm{PCC}(GT_0,GT_1)\)}} \\ 
\hline
10166 & 15.7  & 9.1 & 0.986 & 0.974 & 5.1 & 6.8 & 6.2 & 8.1 & 4.4 & 4.8 & 0.992 \\
\hline
10786 & 32.6 & 29.5 & 0.957 & 0.925 & 3.8 & 4.3 & 3.4 & 7.6 & 3.1 & 3.5 & 0.974 \\
\hline
10280 & 17.8 & 14.9 & 0.981& 0.957& 3.9& 4.4& 4.3& 7.0& 3.4& 3.8& 0.991\\
\hline
11120 & 44.7 & 41.1 & 0.989 & 0.863 & 4.1 & 4.6 & 6.0 & 8.3 & 3.2 & 3.7 & 0.965 \\
\hline
10409 & 45.3 & 39.2 & 0.960 & 0.884 & 3.5 & 4.0 & 4.0 & 8.3 & 3.0 & 3.3 & 0.988\\
\hline
10374 & 35.0 & 24.8 & 0.991& 0.969& 3.7 & 4.1& 4.1& 6.5& 3.0 & 3.5& 0.996\\
\hline
10399 & 25.5 & 21.6 & 0.992& 0.917& 3.7 & 4.1& 4.0& 6.1& 3.1 & 3.4& 0.996\\
\hline
10648 & 13.3 & 10.6 & 0.982& 0.934& 3.8& 4.1& 4.3& 6.5& 3.4& 3.6& 0.994\\
\hline
\hline
\makecell{simulated \\ 10648} & 6.0 & NA & 0.996 & 0.935& 4.5 & 4.6 &  4.6 & 4.8 & 3.5& 4.6& 0.998\\
\hline
\makecell{consensus \\ 10648} & 8.3 & 8.1 & 0.971& 0.893&3.8& 4.1& 4.2& 6.8& 3.4& 3.6& 0.986\\
\hline

\end{tabular}

\begin{itemize}
    \item \(\mathrm{MAnE}\): Mean Angular Error; \(\mathrm{wMAnE}\): weighted Mean Angular Error; \(\mathrm{PCC(V_0,V_1)}\): Reconstructed half-to-half Pearsons’s Correlation Coefficient; \(\mathrm{PCC}(GT, V)\): Reconstructed to ground truth Pearson’s Correlation Coefficient; \(\mathrm{FSCR_{0.143}(V_0,V_1)}\): Reconstructed half-to-half FSC resolution at threshold 0.143, and \(\mathrm{FSCR_{0.5}(V_0,V_1)}\) at threshold 0.5; \(\mathrm{FSCR_{0.143}(GT, V)}\): Reconstructed to ground truth resolution at threshold 0.143, and \(\mathrm{FSCR_{0.5}}(GT, V)\) at threshold 0.5; \(\mathrm{FSCR_{0.143}(GT_0,GT_1)}\): Ground truth half-to-half FSC resolution at threshold 0.143, and \(\mathrm{FSCR_{0.5}(GT_0,GT_1)}\) at threshold 0.5. \(\mathrm{PCC}(GT_0,GT_1)\): Ground truth half-to-half Pearsons’s Correlation Coefficient.
    \item All reported resolutions were obtained using manually computed masks that are available at https://zenodo.org/record/8392782.
\end{itemize}
\end{table}

\newpage

\section{Reconstructed volumes} \label{appx:Reconstructed_volumes}
This appendix shows the volumes reconstructed for some of the best performing examples of the Image2Sphere model on our benchmark. As is shown in all the cases, the quality of the central region of the protein is quite close to the one of the ground truth. However, the density for the regions that are at the edges of the macromolecule is much worse. This is in line to what could be expected if there were some degree of inaccuracy in the angular estimation, as the magnitude of the errors in the volume is proportional to both the angular error and the radius of the macromolecule. 
\begin{figure}[H]
    \centering
    \includegraphics[width=.9\linewidth]{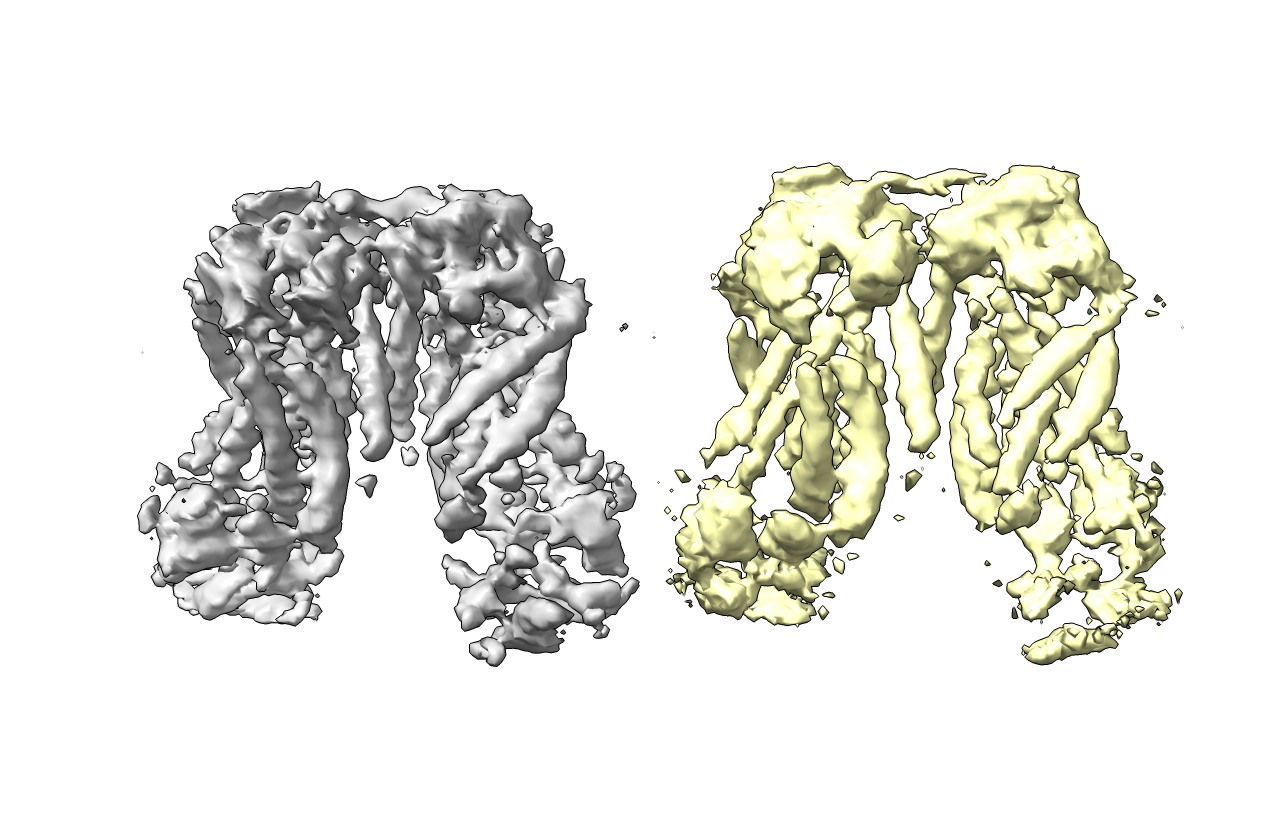}
    \captionsetup{name=Supplementary Figure, labelsep=period} 
    \caption{
Ground truth reconstruction for EMPIAR-10280 (grey) and reconstruction using the angles predicted with the Image2Sphere model (yellow).
}
    \label{fig:empiar-10280-vol}
\end{figure}

\begin{figure}[H]
    \centering
    \includegraphics[width=.9\linewidth]{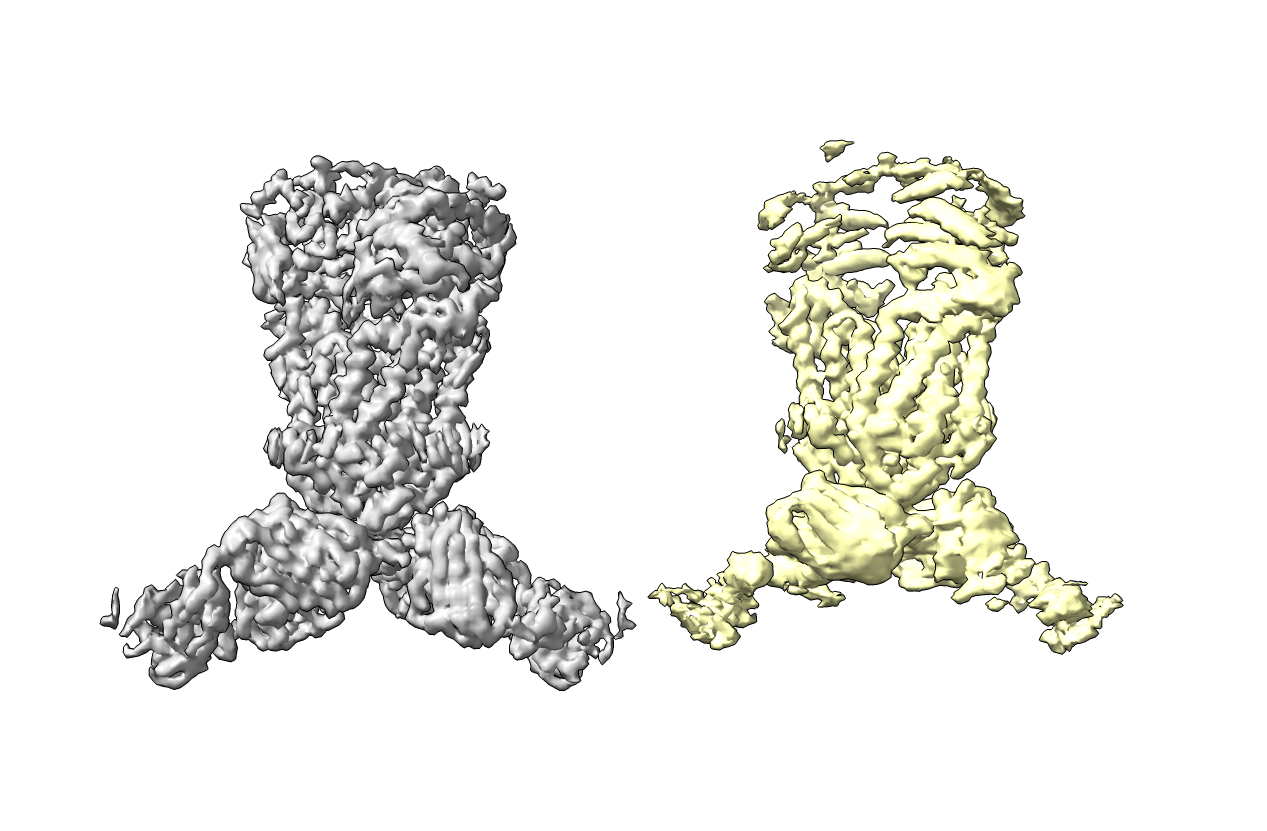}
    \captionsetup{name=Supplementary Figure, labelsep=period} 
    \caption{Groundtruth reconstruction for EMPIAR-10374 (grey) and reconstruction using the angles predicted with the Image2Sphere model (yellow).}
    \label{fig:empiar-10374-vol}
\end{figure}

\section{Locally refined solution}\label{appx:Local_refinement}
In this section, we illustrate the usefulness of our approach by showing the effect of classical local refinement on the Image2Sphere results for the benchmark entry 10374. In this case, the Image2Sphere model predicted poses with a \(\mathrm{wMAnE}\) of 24.8° that lead to a reconstructed map with \(\mathrm{FSCR_{0.143}}(V_0,V_1)\) of 3.7 Å.
When the predicted poses are used as priors for a local refinement in Relion with \verb|--sigma_angle 2.0|, the refined map achieved a \(\mathrm{PCC}(GT, V)\) of 0.997 compared to the original 0.969, showing that the refined map is much more similar to the ground truth map. Indeed, as it can be appreciated in Supplementary Figure \ref{fig:refined-empiar-10786-vol}, after the local refinement, not only the quality of the core of the protein is comparable to the quality of the ground truth, but also the quality of the distant parts of the maps is much better, almost as good as in the ground truth. Equally important, since we are limiting the angular search to the neighbourhood around the predicted poses (\(\pm\) 6°), the number of image comparisons carried out by Relion is much smaller, resulting in a three-fold speed-up in computational time, even when including the time required for pose inference using the Image2Sphere model.

\begin{figure}[ht]
    \centering
    \includegraphics[width=0.9\linewidth]{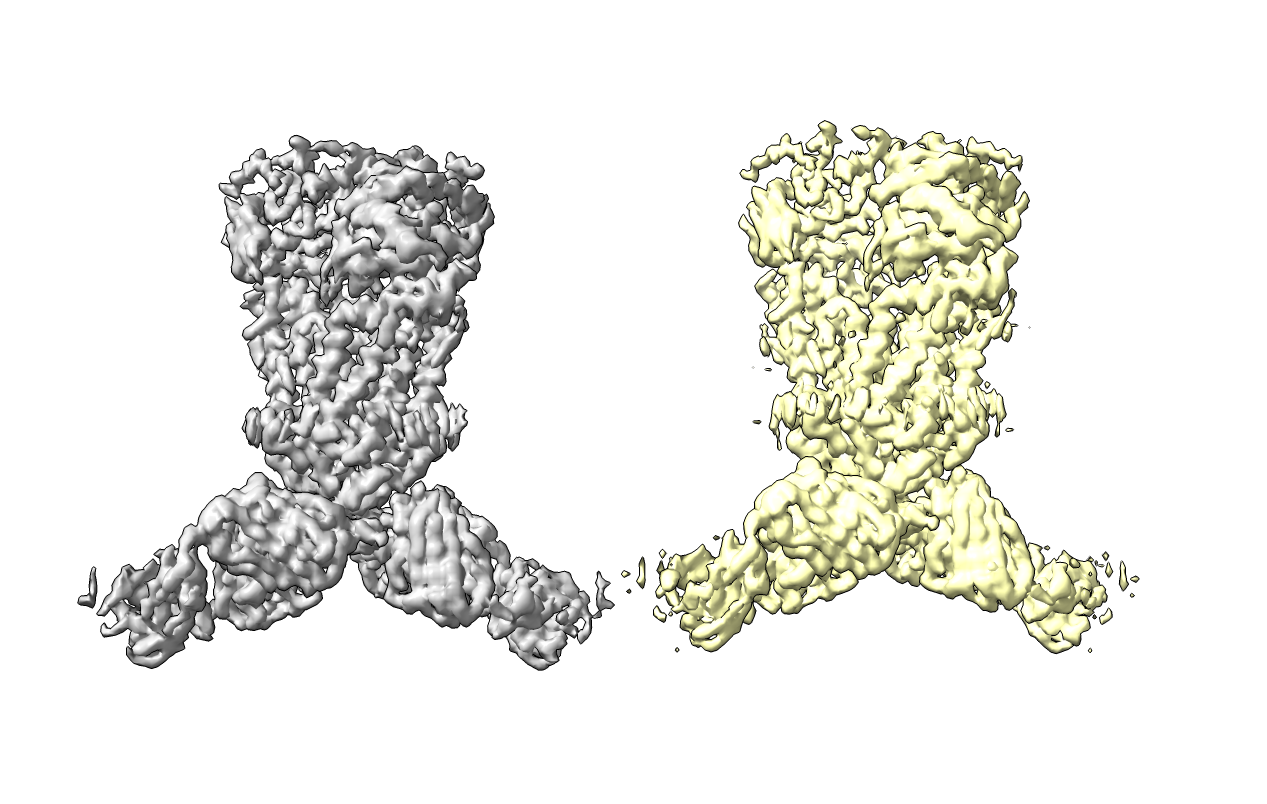}
    \captionsetup{name=Supplementary Figure, labelsep=period} 
    \caption{Ground truth reconstruction for EMPIAR-10374 (grey) and reconstruction using the angles predicted with the Image2Sphere model and locally refined using Relion with priors (yellow).}
    \label{fig:refined-empiar-10786-vol}
\end{figure}
\newpage

\section{Running times} \label{appx:Running_times}
The next table collects the running times of Relion autorefine and the Image2Sphere inference using the same hardware configuration (4 Nvidia A100 cards and 32 CPU cores).
\begin{table}[ht]  
\captionsetup{name=Supplementary Table, labelsep=period} 
\caption{Running time for Relion Autorefine and the Image2Sphere model on CESPED benchmark entries.}  
\small  
\centering  
\begin{tabular}{| l | l | l |}
\hline
\rotatebox{90}{\textbf{EMPIAR ID}} & \textbf{Relion (min)} & \textbf{Image2Sphere (min)}  \\
\hline
10166 & 521 & 22\\
\hline
10786 & 227 & 12 \\
\hline
10280 & 192 & 8 \\
\hline
11120 & 102 &  3 \\
\hline
10648 & 91 & 5 \\
\hline
10409 & 190 &  5 \\
\hline
10374 & 133 & 4 \\
\hline
\end{tabular}

\end{table}

\newpage

\section{Impact of Particle Misalignment in Map Quality Estimation} \label{appx:map_quality_vs_misalignment}

\begin{figure}[ht]
    \centering
    \includegraphics[width=1.\linewidth]{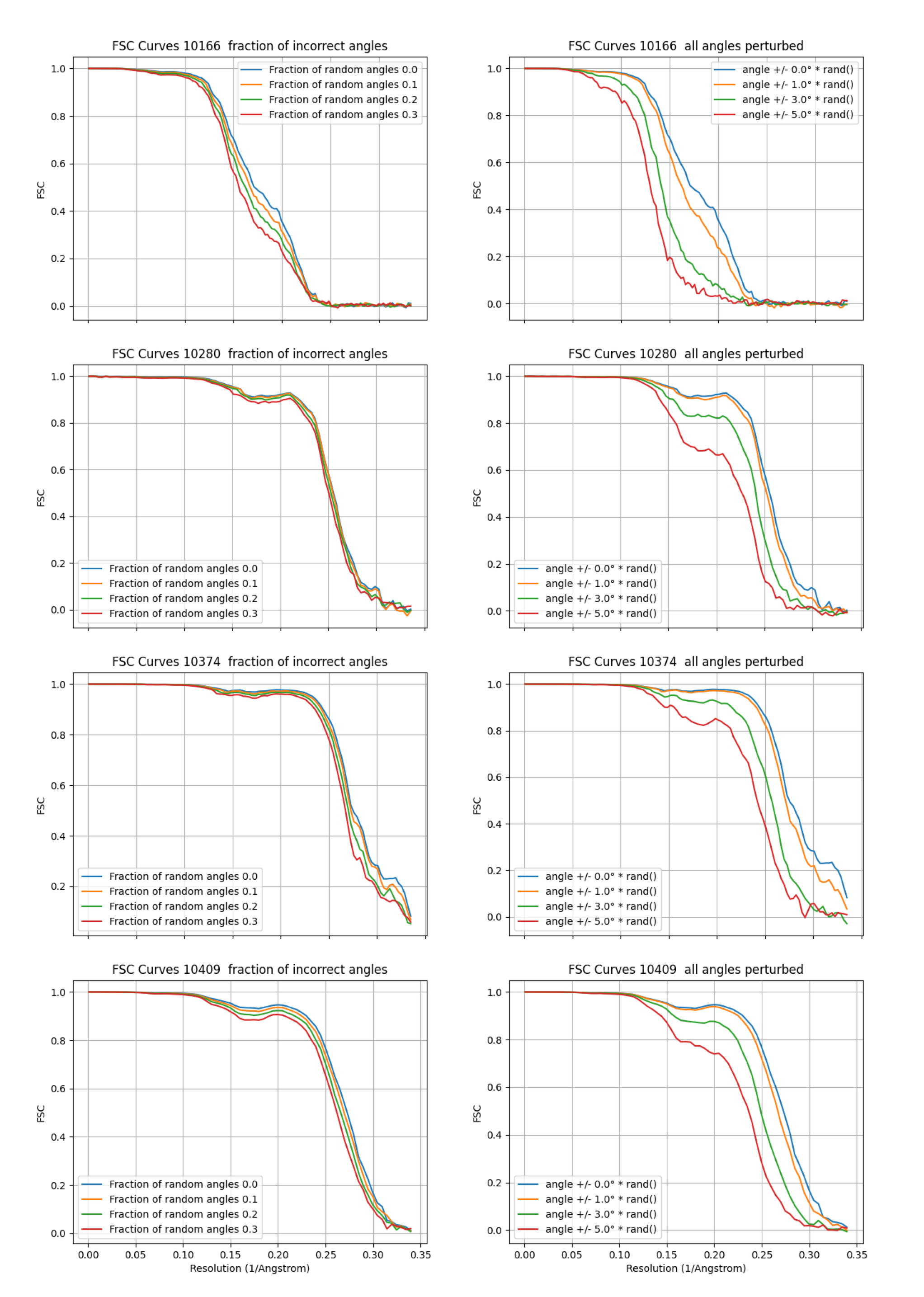}
    \captionsetup{name=Supplementary Figure, labelsep=period} 
    \caption{Effect of particle misalignment in FSC resolution for entries 10166, 10280, 10374, and 10409. Left, FSC curves at different amounts of incorrectly aligned particles: 0\% (blue), 10\% (orange), 20\% (green), and 30\% (red) of the particles were assigned random angles. Right,  FSC curves at different levels of induced angular inaccuracy. Each particle alignment was perturbed using uniform random noise of \(\pm\)0º (blue), 1º (orange),  3º (green), and 5º  (red).}
    \label{fig:fsc_vs_misalignment1}
\end{figure}

\begin{figure}[ht]
    \centering
    \includegraphics[width=1.\linewidth]{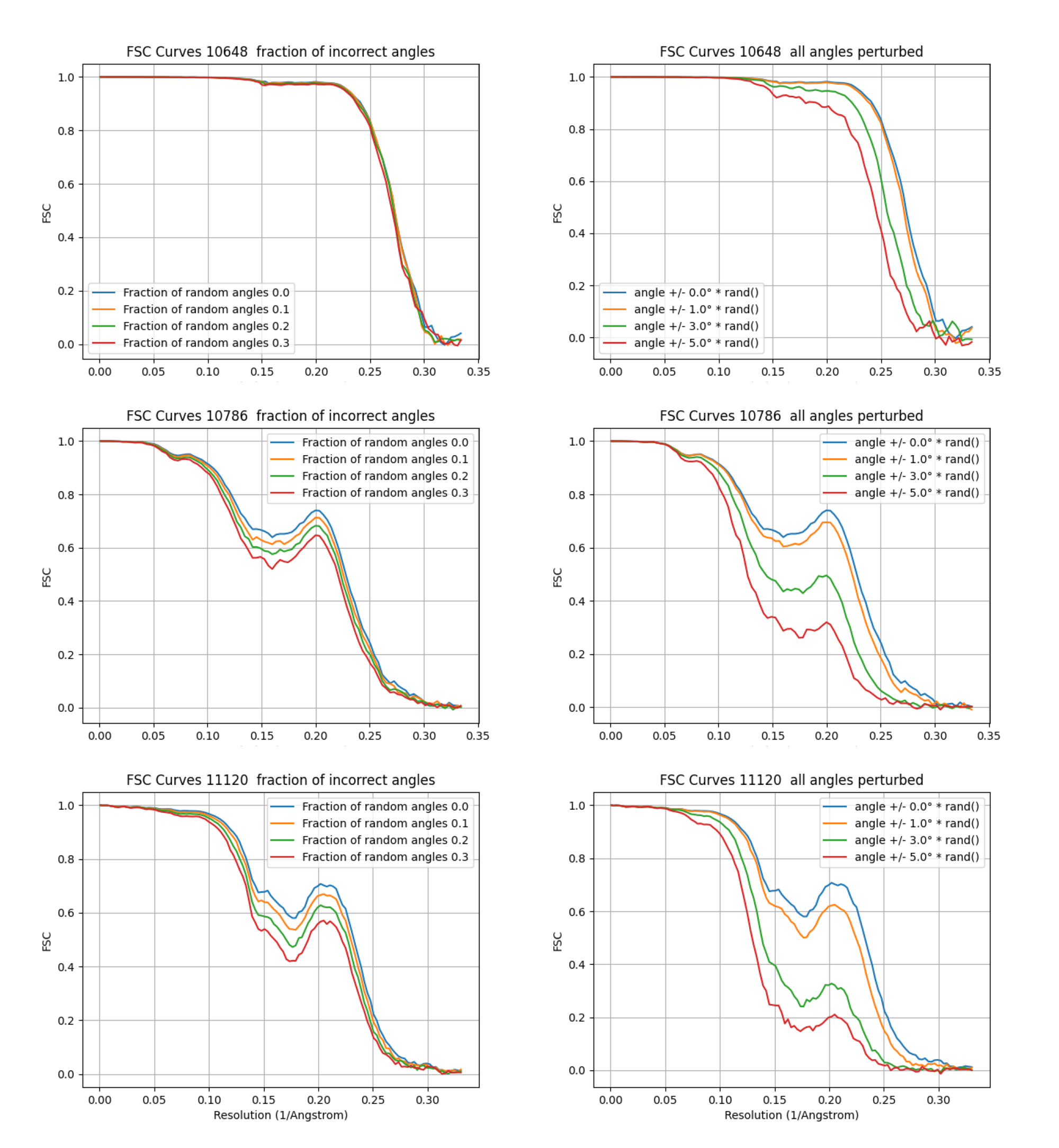}
    \captionsetup{name=Supplementary Figure, labelsep=period} 
    \caption{Effect of particle misalignment in FSC resolution for entries 10648, 10786, and 11120. Left, FSC curves at different amounts of incorrectly aligned particles: 0\% (blue), 10\% (orange), 20\% (green), and 30\% (red) of the particles were assigned random angles. Right,  FSC curves at different levels of induced angular inaccuracy. Each particle alignment was perturbed using uniform random noise of \(\pm\)0º (blue), 1º (orange),  3º (green), and 5º  (red).}
    \label{fig:fsc_vs_misalignment2}
\end{figure}

In order to study the sensitivity to angular inaccuracy of the map quality estimations used in this work, namely the FSC resolution and Pearson's Correlation Coefficient (PCC), we ran two experiments. In the first experiment, we added uniform random noise within the ranges of \(\pm\) 1°, 3°, and 5° to the Euler angles of each particle (Supplementary Table \ref{Suptable_pcc_add_noise} and Supplementary Figures \ref{fig:fsc_vs_misalignment1} and \ref{fig:fsc_vs_misalignment2} right column). In the second experiment we randomised the Euler angles of 10\%, 20\%, and 30\% of the particles for each entry in the benchmark (Supplementary Table \ref{Suptable_pcc_fraction_randomized} and Supplementary Figures \ref{fig:fsc_vs_misalignment1} and \ref{fig:fsc_vs_misalignment2} left column). In the absence of symmetry, the expected angular error (geodesic distance) for randomized angles is approximately ~126.9°, whereas for the uniform random noise, the expected angular error is of ~1.0°, ~2.9°, and ~4.8° respectively (as estimated through simulation). \par

Supplementary Table \ref{Suptable_pcc_add_noise} and the right column of the Supplementary Figures \ref{fig:fsc_vs_misalignment1} and \ref{fig:fsc_vs_misalignment2} illustrate a clear trend in which increasing angular errors lead to a reduction in the FSC resolution and PCC. Since in this experiment we corrupted the alignment of all particles, this underscores that map global quality measurements are effective proxies for estimating overall mean angular accuracy. \par

Supplementary Table \ref{Suptable_pcc_fraction_randomized} and the left column of the Supplementary Figures \ref{fig:fsc_vs_misalignment1} and \ref{fig:fsc_vs_misalignment2}show that as the fraction of misaligned particles increases, both the resolution and the correlation of the maps decreases as well. While it is true that the effect of this type of corruption is smaller than when the angles of all particles are perturbed, it remains noticeable. In most cases, the FSC resolution at threshold 0.5 is clearly different even when as little as 10\% of the particles are perturbed. Given that the number of misaligned particles in refined maps using methods such as Relion is quite large, with some cases reporting up to 60\% misalignment levels\textsuperscript{2}, the sensitivity of the FSC resolution should be enough to compare the accuracy of different methods.  A similar trend is observed in the PCC values, which steadily decline as the fraction of misaligned particles increases.\par

These two experiments confirm that it is possible to distinguish between different levels of alignment corruption using map quality measurements; hence, they serve as sensible proxies for assessing angular alignment accuracy. However, these measurements are not directly comparable across different samples; thus, comparisons are only valid when examining different alignment results for the same sample, as we do in this benchmark. \par
\begin{table}[H]
\centering
\captionsetup{name=Supplementary Table, labelsep=period} 
\centering
\caption{Volume Pearson's Correlation Coefficients (PCC) at different levels of induced angular inaccuracy. For each particle in each entry, uniform random  noise of \(\pm\)0º, 1º, 3º, and 5º is added to all the components of the Euler angles. PCC values are reported without using a mask, whereas Masked PCC values are computed using the mask described in the manuscript. 
}
\begin{tabular}{lcccc|cccc}
\hline
\textbf{Entry} & \multicolumn{4}{c}{\textbf{PCC}} & \multicolumn{4}{c}{\textbf{Masked PCC}} \\
 & \textbf{0°} & \textbf{1º} & \textbf{3º} & \textbf{5º} & \textbf{0º} & \textbf{1º} & \textbf{3º} & \textbf{5º} \\
\hline
\label{Suptable_pcc_add_noise}
10166 & 0.981 & 0.950 & 0.921 & 0.894 & 0.995 & 0.989 & 0.968 & 0.941 \\
10280 & 0.978 & 0.946 & 0.911 & 0.887 & 0.995 & 0.991 & 0.972 & 0.949 \\
10374 & 0.980 & 0.950 & 0.925 & 0.908 & 0.998 & 0.996 & 0.980 & 0.962 \\
10409 & 0.960 & 0.914 & 0.833 & 0.781 & 0.976 & 0.951 & 0.886 & 0.835 \\
10648 & 0.966 & 0.916 & 0.868 & 0.821 & 0.997 & 0.995 & 0.969 & 0.917 \\
10786 & 0.931 & 0.840 & 0.749 & 0.700 & 0.938 & 0.858 & 0.772 & 0.722 \\
11120 & 0.852 & 0.571 & 0.433 & 0.391 & 0.905 & 0.736 & 0.606 & 0.555 \\
\hline
\end{tabular}
\end{table}

\begin{table}[H]
\captionsetup{name=Supplementary Table, labelsep=period} 
\centering
\caption{Volume Pearson's Correlation Coefficients (PCC) at different amounts of incorrectly aligned particles. For each entry, 0\%, 10\%, 20\%, and 30\% of the particles were assigned random angles. PCC values are reported without using a mask, whereas Masked PCC values are computed using the mask described in the manuscript. 
}
\begin{tabular}{lcccc|cccc}
\hline
\textbf{Entry}& \multicolumn{4}{c}{\textbf{PCC}} & \multicolumn{4}{c}{\textbf{Masked PCC}} \\
 & \textbf{0\%}& \textbf{10\%}& \textbf{20\%}& \textbf{30\%}& \textbf{0\%}& \textbf{10\%}& \textbf{20\%}& \textbf{30\%}\\
\hline
\label{Suptable_pcc_fraction_randomized}
10166 & 0.981 & 0.971 & 0.956 & 0.935 & 0.995 & 0.992 & 0.985 & 0.975 \\
10280 & 0.978 & 0.964 & 0.946 & 0.926 & 0.995 & 0.988 & 0.975 & 0.958 \\
10374 & 0.980 & 0.971 & 0.959 & 0.945 & 0.998 & 0.995 & 0.987 & 0.976 \\
10409 & 0.960 & 0.944 & 0.924 & 0.899 & 0.976 & 0.965 & 0.949 & 0.928 \\
10648 & 0.966 & 0.941 & 0.909 & 0.875 & 0.997 & 0.988 & 0.970 & 0.949 \\
10786 & 0.931 & 0.904 & 0.874 & 0.838 & 0.938 & 0.914 & 0.886 & 0.853 \\
11120 & 0.852 & 0.805 & 0.753 & 0.699 & 0.905 & 0.872 & 0.834 & 0.792 \\
\hline
\end{tabular}
\end{table}

\newpage
\newpage

\end{document}